\documentclass[prl,a4paper,twocolumn,superscriptaddress,showkeys]{revtex4}

\usepackage{graphicx}
\usepackage{amsmath}
\usepackage{bm}
\usepackage{xspace}

\begin{document}

\title{Hot Charge Transfer States and Charge Generation in Donor Acceptor Blends}

\author{James Kirkpatrick}
\email{james.kirkpatrick@ic.ac.uk}
\affiliation{Department of Physics, Imperial College London, Prince Consort Road, London SW7 2BW, UK}

\date{\today}

\begin{abstract} 
In an organic blend the vibrational normal mode excited by exciton splitting is
the same as the one coupled to charge hopping. Excess driving force for exciton splitting can 
therefore aid charge transfer, if vibrational relaxation is slow compared to 
charge transfer. A model is developed that takes this into account and hence explains 
the experimentally observed relation of driving force for exciton splitting and 
charge yield and that high charge yields can be achieved with the
observed fast rates of recombination.
\end{abstract}


\maketitle 

\section{Introduction}
In an organic blend, charges are generated by splitting excitons 
at the interface between materials with different electron affinities: the donor
 material is typically responsible for absorbing light and has a  
greater affinity for holes; the acceptor has greater affinity for electrons.  
Charges must escape 
each other's Coulomb attraction without recombining. Exciton splitting 
is typically driven by a large ($>0.5~eV$) driving force \cite{Vandewal2009}. 
We develop a model where this excess energy aids charges escape from the Coulomb 
well. This can happen because the polaron pair formed by exciton splitting is 
in a vibrational excited state. The mode excited is the superposition of the modes 
that change the donor from the exciton to the charged geometry and the acceptor 
from the neutral to the charged geometry. This second mode is 
also coupled to charge transfer in the acceptor molecule, hence, 
if the mode is sufficiently long lived, the states formed just after exciton splitting 
are able to separate more efficiently if extra energy is available.

That hot states are involved in charge separation is not a new idea.
Peumans \emph{et al.} need to assume very large thermalization radii 
to reproduce experimental charge yields \cite{Peumans2004}. Models involving 
hot excitons have been used to describe charge generation in pure 
polymers \cite{Arkhipov1999}. Models by Offermans \emph{et al.} 
predict increased charge yields if charges are generated hot \cite{offermans2003}. 
Experiments have also shown that charge mobilities are higher immediately 
after charge generation \cite{
		Cabanillas-Gonzalez2007}, suggesting that 
charge transfer is fastest immediately after exciton splitting.
However, \emph{how} thermalization 
lengths are connected to molecular properties is not yet understood.

Understanding the link between the driving force for charge separation and quantum 
yield will help determine the limiting efficiency of organic solar cells (OSC). OSCs can have
high quantum efficiencies \cite{Park2009}, however open circuit voltages (Voc) 
are still substantially smaller than the band gap.
The Voc for several materials is less than half the maximum of external quantum 
efficiencies  \cite{Vandewal2009}.
Until recently, designing better OSC materials was 
thought to require reducing the energy lost in exciton splitting \cite{Scharber2006}. 
Recent measurements by Okhita \emph{et al.} \cite{Ohkita2008}
relate the energy loss and the yield of charges: 
bad news for OSC as increasing the Voc would also lower quantum yields.

Modeling generation without thermalization requires recombination rates in the 
microsecond range \cite{offermans2003, Offermans2005, Groves2008,casalegno:094705} 
in order to achieve high yields of charges ($> 10\%$). But
experimental measurements of the rate of charge recombination have found it to be faster 
than nanoseconds \cite{Montanari2002, Veldman2008}. Here 
high yields are obtained with fast recombination by including the effect of charge 
transfer from hot states.
 

The theoretical background for the model developed is given by the 
Marcus Levitch Jortner (MLJ) equation \cite{Walker1992}. 
The rate of charge transfer becomes:
\begin{equation}
\label{eq:MLJ}
\Gamma = \frac{\langle i |H|j\rangle^2}{\hbar} \sqrt{\frac{\pi}{\lambda_o kT}} \Sigma_f \langle 0|f\rangle^2 
e^{ \left(-\frac{(\Delta G + f \hbar \omega + \lambda_o )^2 }{4 \lambda_o kT}\right)}
\end{equation}
where $\langle i|H|j\rangle$ is the matrix element for charge tunneling 
between the diabatic electronic states $i$ and $j$, $\lambda_o$ is the 
outer sphere reorganization energy four coupling to a classical
bath, $\omega$ is the frequency of the 
quantum mode and $\langle 0|f\rangle$ is the Franck Condon overlap between
 the initial vibrational state (the ground state) and the final one (with quantum number f).
 The Franck Condon factor is computed assuming that the initial and final 
vibrational states are wavefunction of simple harmonic oscillators with identical 
frequencies but shifted in mass weighted coordinates by $\sqrt{\Delta}$, 
the Huang Rhys factor. $\Delta G $ represents the difference in free energy between the 
initial and final diabatic states.  Each term in the summation represents the 
probability of resonant electron tunneling occurring between the ground vibrational state 
of the initial state and the $f^{th}$ vibrational state of the final states. 
This equation can be used to model all three processes of interest in the paper: 
exciton splitting, charge transfer and charge recombination. 
A caveat is that charge transfer must occur in the non-adiabatic regime, 
\emph{i.e.} the transfer integral must be small compared to the electron 
$\lambda_o $and to  $\Delta \hbar \omega$. 

Figure \ref{fig:start} shows the processes modeled in the paper. Each column of 
energy levels represents an electronic state, labeled by the distance $d$ between 
electron and hole. At a distance 0 the system is in the neutral ground state 
(black energy levels) or in the excited state (red energy level). 
The processes competing are: i) exciton dissociation; ii) charge transfer from a 
vibrationally excited state; iii) decay of the vibrational energy by internal conversion; 
and iv) charge recombination to the ground state. Exciton splitting is not modeled 
explicitly, but is used to provide the initial excited vibrational state.


\section{Method}

The vibronic state $ |i, \bf{\nu}\rangle$ of the system is described by 
the outer product of the electronic state $|i\rangle$ and of the vector 
describing the vibrational state of the oscillators $|\bf{\nu}\rangle$. 
Each mode in $|\bf{\nu}\rangle$ is coupled to the energy of a single 
electronic state $|i\rangle$.

When charge transfer occurs between two electronic states $i$ and $j$ , 
only the modes of the oscillators coupled to each of those two states 
will be able to change.
If the mode associated with electronic state $i$ changes from quantum number 
$\mu$ to $\rho$ and the mode associated to $j$ changes from
$\nu$ to $\sigma$, the rate of charge transfer can be written as:
\begin{eqnarray}
\Gamma &(&i,\mu,\nu \rightarrow  j, \rho, \sigma) = \\ \nonumber
& &  \sum_{S',S} P(S,A |\nu, \mu) \Gamma(i,S \rightarrow j, S') P(S',A | \rho, \sigma)
\end{eqnarray}
where $\Gamma(i,S \rightarrow j, S') $ represents the rate of transfer between 
electronic states $i$ and $j$ and symmetric combination of the vibrational modes 
$S$ to $S'$. Only symmetric combinations
of modes contribute to charge transfer, as the energy difference between two 
oscillators depends only
on the difference of the distorsions of each oscillator.
The probability terms $P(S,A|\mu, \nu)$ represent the probability of a system of two simple harmonic 
oscillators in localised modes $\mu$ and $\nu$ to be in a symmetric
 mode $S$ and an antisymmetric
mode $A$. By using a probability term, rather than explicitly considering
 the wavefunction, we
implicitly assume that with each charge hop, coherence is lost.

The rate term $ \Gamma(i,S \rightarrow j, S') $ can be computed with a 
trivial generalisation of the MLJ equation:
\begin{eqnarray}
\label{eq:avRate}
 \Gamma(i,S \rightarrow j,S') & = &  \frac{\langle i |H|j\rangle^2}{\hbar} \sqrt{\frac{\pi}{\lambda_o kT}} \langle S|S'\rangle^2  \\ \nonumber
 & & exp \left(-\frac{(\Delta G_{i,j} + \lambda_o + \hbar \omega( S' - S))^2 }{4 \lambda_o kT}\right)
\end{eqnarray}
where all symbols have the same meaning as in equation 1. 

The probabilities $P(S,A | \mu, \nu)$ can be determined by assuming that 
the ladder operators for the symmetric/antisymmetric modes 
are given by combinations of the appropriate localised ladder 
operators $a^\dagger_1$ and $a^\dagger_2$: 
$ a_{s/a}^\dagger  =  \frac{a_1^\dagger \pm a_2^\dagger}{\sqrt{2}} $ 
The amplitude of the projection of such a state into the corresponding 
localised modes  is:
\begin{eqnarray}
\label{eq:ProbVib}
&P&(S,A| \mu, \nu)  =  2^{-(\mu+\nu)} \\ \nonumber
& &\Bigg(   \sum_{\substack{ 0 \le m \le \mu \\ 0 \le n \le \nu}}  \begin{array}{c} \mu \\ m \end{array} 
 \begin{array}{c} \nu \\ n \end{array} \sqrt{\frac{(m+n)!(\mu+\nu-m-n) !}{\mu ! \nu !}}  \\  \nonumber
& & \delta(A,(m+n)) \delta( S,(\mu+\nu-m-n)) \Bigg)^2 
\end{eqnarray}
where the
$\delta$ functions 
ensures that 
the probability is non-zero only 
if $\mu+\nu= A+S$. 

As well as charge hopping between vibronic states, a state is also 
allowed to lose a vibrational quantum by internal conversion
at a rate $k_{VR}$ without changing its electronic state.

\begin{figure}
\centering
\includegraphics[width=8cm]{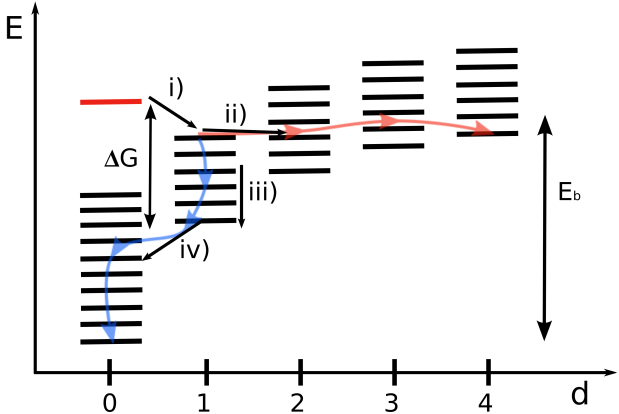}
\caption{ Sketch of the mechanisms described in the main text: i) exciton splitting, ii) charge transfer from a vibrational excited state, iii) internal vibrational relaxation, iv) recombination. The red arrow is a trajectory leading to charge generation whereas the blue one to recombination. The pseudo bandgap $E_b$ and the driving energy for exciton splitting $\Delta G$ are shown. }
 \label{fig:start}
\end{figure}

The dynamical system defined by these equations is linear and could be solved by 
writing a master equation, but since the total vibronic phase space is large 
it is more efficient to solve it by generating a starting state vector 
$ |1, \bf{\nu}\rangle$ and updating it using a continuous time random walk 
algorithm \cite{Klafter80}. Rates are then computed only for states 
as they are needed. The simulation is stopped if either the charges recombines, 
or if they reaches a certain distance $d$.  We do not explicitly 
model exciton dissociation, but use the parameters for exciton dissociation 
to determine the probability of starting the simulation from a 
particular vibrational level, given by:
\begin{equation}
P(\nu) = \sum_S \frac {\Gamma(0^*, 0 \rightarrow 1, S)}{\sum_{S'}{\Gamma(0^*, 0 \rightarrow 1, S') } } P(S,0 | S-\nu, \nu)
\end{equation}
where each term in the summation represents the probability of producing a certain 
symmetric state $S$ and the probability that the state gives rise to 
localised state $\nu$. The excited neutral state is labeled $0^*$. 
Note that this expression depends on the difference in energy between the 
excited state and the first charged state (labelled $\Delta G$ in figure \ref{fig:start}), 
on $\lambda_o$, on the Huang Rhys factor and on $\hbar \omega$, but not on the matrix
element.

We apply this model it to a one dimensional chain of acceptors, with the hole localised at 
one end of the chain. Electronic states are uniquely labeled by the distance 
between hole and electron $|i\rangle$, for example state $|1\rangle$ will have 
the electron and hole one lattice apart spacing. State $|0\rangle$ is the 
neutral ground state reached if the electron and hole recombine.  
Each state $|i\rangle$ is connected to its neighbors $|i+1\rangle$ and $|i-1\rangle$ with 
the same transfer integral for charge separation $V_{CS}$, with the 
exception of states $|1\rangle$ and $|0\rangle$, which are 
connected by the transfer integral for charge recombination $V_{CR}$, 
this allows the timescales for charge transfer and recombination to be controlled 
independently.

The energy $E(i)$ of a particular electronic state $|i\rangle$ is determined 
only by the applied electric field $F$ and the Coulomb potential: $E(i) = - \frac{e^2}{4 \pi \epsilon_0 \epsilon ~  a ~ i} - F ~  a ~ i
$, where $a$ is the lattice constant, $\epsilon$ and $\epsilon_0$ are respectively the 
relative and vacuum permittivities and $e$ is the charge of an electron. 
The energy of state $|0\rangle$  is a fixed value $E_b$ below the zero of this potential. 
$E_b$ is the the pseudo band gap between the 
highest occupied molecular orbital of the donor and the lowest 
unoccupied molecular orbital of the acceptor. 
$E_b$ and $\Delta G$ (the driving force for exciton splitting) are shown schematically in
 figure \ref{fig:start}. 

\section{Results}

All our modeling uses  the following  
parameters for the energetics: 
$a=1~nm$, $E_b = -0.9~eV$, $\epsilon=4$ and $F=5~10^5 V/cm$. 
The electric field employed is rather large, smaller fields would be necessary 
if more subtle models of the polarization of the interface 
\cite{Linares2010, Szmytkowski2009} lowered the barrier do charge separation, or if
a three dimensional model which allowed both charges 
to move was employed \cite{Offermans2005, casalegno:094705}. 

\begin{figure*}
\centering
\includegraphics[width=12cm]{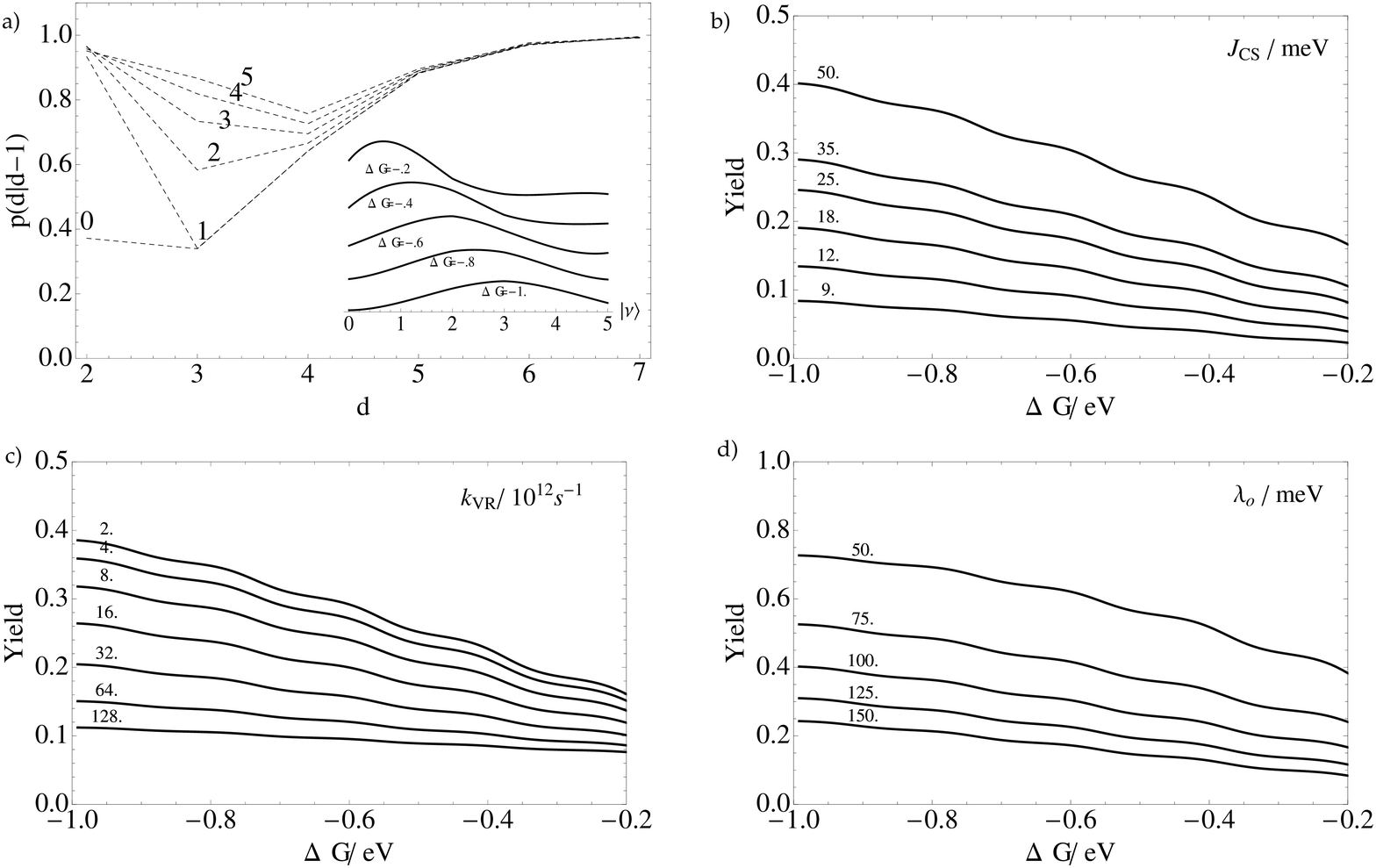}
\caption{Panel a) The conditional probability of the system reaching a certain distance $d$ given that it has reached distance $d-1$ for different initial quanta of vibrational excitation. The inset shows the probability distribution of producing a certain number of quanta for various values of the driving force for exciton splitting. The curves are offset for clarity. The value of $\Delta G$ in eV is shown below each curve. Panels b), c) and d) show the yield of charges at 5 nm as a function of the following parameters: b) the transfer integral for charge separation, c) the rate of internal relaxation, d) the outer sphere reorganization energy. Each curve is labelled by the value of the parameter used in that simulation.}
\label{fig:results}
\end{figure*}

The parameters of the charge transfer equation used are:
  $\Delta=1$, $\hbar \omega =0.17 ~ eV$, 
$\lambda_o = 0.1~eV$, $V_{CS}=0.05~eV$, and $k_{VR}=10^{12}~ s^{-1}$, a set of values
 typical for fullerene \cite{Kwiatkowski2009}. The parameters for 
recombination are the same as above, but with a transfer integral $V_{CR}=0.01~eV$ and
 an energy defined by $E_b=-1~eV$. This puts the recombination rate 
in the sub ps regime. Panel a) in figure \ref{fig:results} shows the conditional 
probability of the system reaching a certain distance $d$ given 
that it has already reached a distance $d-1$ for a range of initial 
vibrational quanta. Showing the conditional probability helps identify 
the thermalization distance, because once the system thermalizes all the curves 
fall onto each other. Increasing the number of phonons in the initial state massively 
increases the probability that the charge is able to escape recombination, 
for example in the vibrational ground state the probability of recombining 
immediately is $60\%$, but with just one quantum of vibrational energy this 
probability decreases to less than $1\%$. 
Simulations with one extra phonon thermalise after just one hop. 
The higher excited states thermalise at a distance of 4nm, showing that 
(for these parameters) thermalization is complete after three or four hops. 
The inset of panel a) shows the probability of a particular vibrational excited 
state being created given a certain $\Delta G$ . Clearly 
highly excited states are more likely for greater $\Delta G$. 
Weighing the probabilities computed by the probability that that 
starting number of phonons is achieved for a certain $\Delta G$ gives the 
charge yield as a function of $\Delta G$. 

Panels b),c) and d) show the effect on charge yield at $5 ~nm$ as a function of 
$\Delta G$ of: the transfer integrals 
for charge transport (b), the rate of vibrational relaxation (c) and 
$\lambda_o$ for charge transport (d). 
Loss of vibrational excitation can occur in two ways: by internal vibrational 
relaxation or by losing energy to the classical modes through 
the external reorganization energy. The yield of charges is therefore 
the result of competition of three basic process: charge transport, 
internal vibrational relaxation and charge recombination.
Decreasing the transfer integral not only reduces the yield of charges, 
it also decreases the dependence of yield on $\Delta G$. This is 
because slowing down charge transfer not only makes recombination more 
likely (reducing the yield) it also makes vibrational relaxation 
faster (reducing the dependence on $\Delta G$). Increasing the rate 
of vibrational relaxation reduces the yield because it causes 
relaxation to occur faster. $\lambda_o$ has a large effect 
because it dictates how much energy is lost to the classical bath with each hop: 
making it larger makes the system thermalise more effectively through charge transfer.

\section{Conclusion}

Explicitly modeling charge transfer between vibrationally excited
 states allows high charge yields 
with fast recombination and explains the dependence of charge yield on 
the driving force for exciton splitting. Improving charge transport improves 
charge generation  because of more effective competition of 
charge separation with both charge recombination and vibrational relaxation.

Three conditions must be fulfilled for vibrational 
excitation to aid charge generation : 1) the vibrational mode excited by 
exciton splitting must be the same as that coupled to charge separation, 
2) vibrational relaxation must be slow compared to charge tunneling, 
3) $\lambda_o$ must be small 
to ensure low losses to the classical bath. 



The present model makes two fundamental by two assumptions: that coherence 
is lost at each hop and that the coupling is non-adiabatic. 
Future work concentrate on tackling these problems. 
The model allows us to predict that the quantum efficiency of 
a blend should increase with photon energy. We 
have assumed that vibrational
excitation aids charge separation
only in the acceptor molecule. 
If the modes that lead to exciton splitting 
are not orthogonal to those coupled to charge transfer also in the donor molecule, 
    the effect of the initial 
energy would be increased. This provides a design rule for good donor materials.


This work was supported by the Engineering and Physical Sciences Research Council. We acknowledge
J. Nelson, A. Horsfield, A. Fisher, L. Stella, R. Miranda and J. Frost for useful discussion.

\end{document}